\documentclass[aps,prb,twocolumn,floatfix,superscriptaddress]{revtex4-1}

\usepackage{amsmath}
\usepackage{amssymb}
\usepackage{times}
\usepackage{braket}
\usepackage{graphicx}
\usepackage{blindtext}
\usepackage{hyperref}
\usepackage{cleveref}
\usepackage{color}

\renewcommand{\vec}[1]{\mbox{\boldmath$\mathrm{#1}$}}
\def\ind#1{{_{\mathrm{#1}}}}

\newcommand{\dd}{\mathrm{d}}

\newcommand{\bcor}{\color{black} }
\newcommand{\bcorr}{\color{black} }

\begin{document}
	
\title{Stochastic dynamics and pattern formation of geometrically confined skyrmions} 

\author{Alexander F. Sch{\"a}ffer}
\email[Corresponding author. ]{alexander.schaeffer@physik.uni-halle.de}
\affiliation{Institut f\"ur Physik, Martin-Luther-Universit\"at Halle-Wittenberg, D-06099 Halle (Saale), Germany}

\author{Levente R\'{o}zsa}
\affiliation{Institut f\"ur Nanostruktur- und Festk\"orperphysik, Universit\"at Hamburg, D-20355 Hamburg, Germany}

\author{Jamal Berakdar}
\affiliation{Institut f\"ur Physik, Martin-Luther-Universit\"at Halle-Wittenberg, D-06099 Halle (Saale), Germany}

\author{Elena Y. Vedmedenko}
\affiliation{Institut f\"ur Nanostruktur- und Festk\"orperphysik, Universit\"at Hamburg, D-20355 Hamburg, Germany}

\author{Roland Wiesendanger}
\affiliation{Institut f\"ur Nanostruktur- und Festk\"orperphysik, Universit\"at Hamburg, D-20355 Hamburg, Germany}

\date{\today}

\begin{abstract}
	Ensembles of magnetic skyrmions in confined geometries are shown to exhibit thermally driven motion on two different time scales. The intrinsic fluctuating dynamics ($t\sim 1~$ps) is governed by short-range symmetric and antisymmetric exchange interactions, whereas the long-time limit ($t\gtrsim10\,$ns) is determined by the coaction of skyrmion-skyrmion-repulsion and the system's geometry.  
    Micromagnetic simulations for realistic island shapes and sizes are performed and analyzed, indicating the special importance of skyrmion dynamics at finite temperatures. 
	We demonstrate how the competition between skyrmion mobility and observation time directly affects the addressability of skyrmionic bits, which is a key challenge on the path of developing skyrmion-based room-temperature applications.
	The presented quasiparticle Monte Carlo approach offers a computationally efficient description of the diffusive motion of skyrmion ensembles in confined geometries{\bcor,} like racetrack memory setups.
\end{abstract}
\maketitle
	Magnetic skyrmions~\cite{bogdanov1989thermodynamically,muhlbauer2009skyrmion,nagaosa2013topological} are quasiparticles which are considered as possible carriers of information for future storage devices. Their specific chirality is determined by the antisymmetric Dzyaloshinskii-Moriya exchange interaction (DMI)~\cite{dzyaloshinsky1958thermodynamic,moriya1960anisotropic}. The DMI can be induced by a broken inversion symmetry in a crystal itself (e.g. MnSi)~\cite{muhlbauer2009skyrmion} or by interfacing a heavy metal layer (e.g. Pt, W, Ir) with a ferromagnetic material (e.g. Fe, Co)~\cite{heinze2011spontaneous,RoKu2015}.
	 
	Conceptually the utilization of these topologically non-trivial~\cite{bogdanov1999stability} quasiparticles in so-called racetrack setups is of great interest ~\cite{parkin2008magnetic,sampaio2013nucleation,tomasello2014strategy}: Skyrmions can be manipulated (written and deleted)~\cite{romming2013writing,schaffer2017ultrafast} and addressed individually~\cite{hanneken2015electrical,gobel2018magnetoelectric,maccariello2018electrical} on a magnetic stripe, allowing a memory device extension from a pure surface density into the third dimension by pushing the quasiparticle-hole-train back and forth, e.g. by applying electrical currents~\cite{slonczewski1996current,iwasaki2013universal}. The low threshold driving current~\cite{jonietz2010spin} along with the small size and high stability of the skyrmions are key features of this concept. 
	
	To connect experimental and theoretical model systems with technological applications, investigating the influence of finite temperatures is of crucial importance. 
	The bits on a racetrack-based memory device need to fulfill two main features, stability against external perturbations and addressability. Both are affected by thermal fluctuations, as shown below. 
	
	The stability of skyrmions was examined in several publications over the last years~\cite{hagemeister2015stability,RoSi2016,lobanov2016mechanism,bessarab2018lifetime,von2017enhanced}.
	R\'{o}zsa et al.~\cite{RoSi2016} investigated theoretically periodic two-dimensional Pd/Fe double-layers on Ir(111) and determined the phase diagram as a function of external field and temperature, which includes field-polarized, skyrmion lattice, spin spiral, fluctuation-disordered and paramagnetic regions.
	Skyrmion lifetimes in the fluctuation-disordered regime were calculated. 
	The lifetimes of isolated skyrmions in racetrack geometries for Pd/Fe/Ir(111) and Co/Pt(111) systems were investigated in Refs.~\cite{bessarab2018lifetime,lobanov2016mechanism}, and different mechanisms were revealed for the collapse of a skyrmion inside the track and at the boundary.
		
	The diffusive motion of skyrmions at finite temperature has also attracted significant research attention lately~\cite{PhysRevB.90.174434,miltat2018brownian,zazvorka2019thermal}. Previous studies concentrated on diffusion in infinite or extended geometries, but a clarification of the role of the sample shape still seems to be missing in the case where the system size becomes comparable to that of the skyrmions. Effects of this kind directly impede the addressability, which is indispensable when using skyrmions for storage technology, e.g. in racetrack memory devices. Since the number of skyrmions during the diffusive motion remains constant, quasiparticle models have been developed for their description in this limit~\cite{PhysRevB.90.174434,miltat2018brownian,PhysRevB.87.214419,PhysRevLett.114.217202}, which are primarily based on the Thiele equation~\cite{PhysRevLett.30.230}. The advantage of such a collective-coordinate description over micromagnetic or atomistic spin dynamics simulations is its significantly lower computation cost.

	Due to the finite temporal resolution, imaging techniques on the atomic length scale like spin-polarized scanning tunneling microscopy (SP-STM)\cite{Wi2009} or magnetic force microscopy (MFM)\cite{martin1987magnetic} cannot access temporally the diffusive motion of the skyrmions. Instead, the time-averaged skyrmion probability distribution is imaged which may well be different from the zero-temperature configuration or a snapshot of a simulation. Here we will discuss the formation, stability and addressability of diffusive skyrmion configurations. Using micromagnetic simulations, it is shown that the complex interplay of the repulsive interaction between the skyrmions~\cite{LiRe2013,rozsa2016skyrmions} along with the confinement effect of the nanoisland and the thermally induced skyrmion diffusion leads to a pattern formation of the skyrmion probability distribution on the nanosecond time scale. A computationally efficient quasiparticle Monte Carlo method is introduced, which is found to yield comparable results to time-averaged micromagnetic simulations regarding the skyrmion probability distribution. 
	{\bcor The simple implementation and high speed of such a method may make it advantageous over micromagnetic simulations when the actual number of skyrmions has to be determined based on a time-averaged experimental image.}

\section*{Results}
\subparagraph{Skyrmion stability.}\label{sec_micromagnetic}

	\begin{figure}[hbt!]
		\includegraphics[width=.9\linewidth]{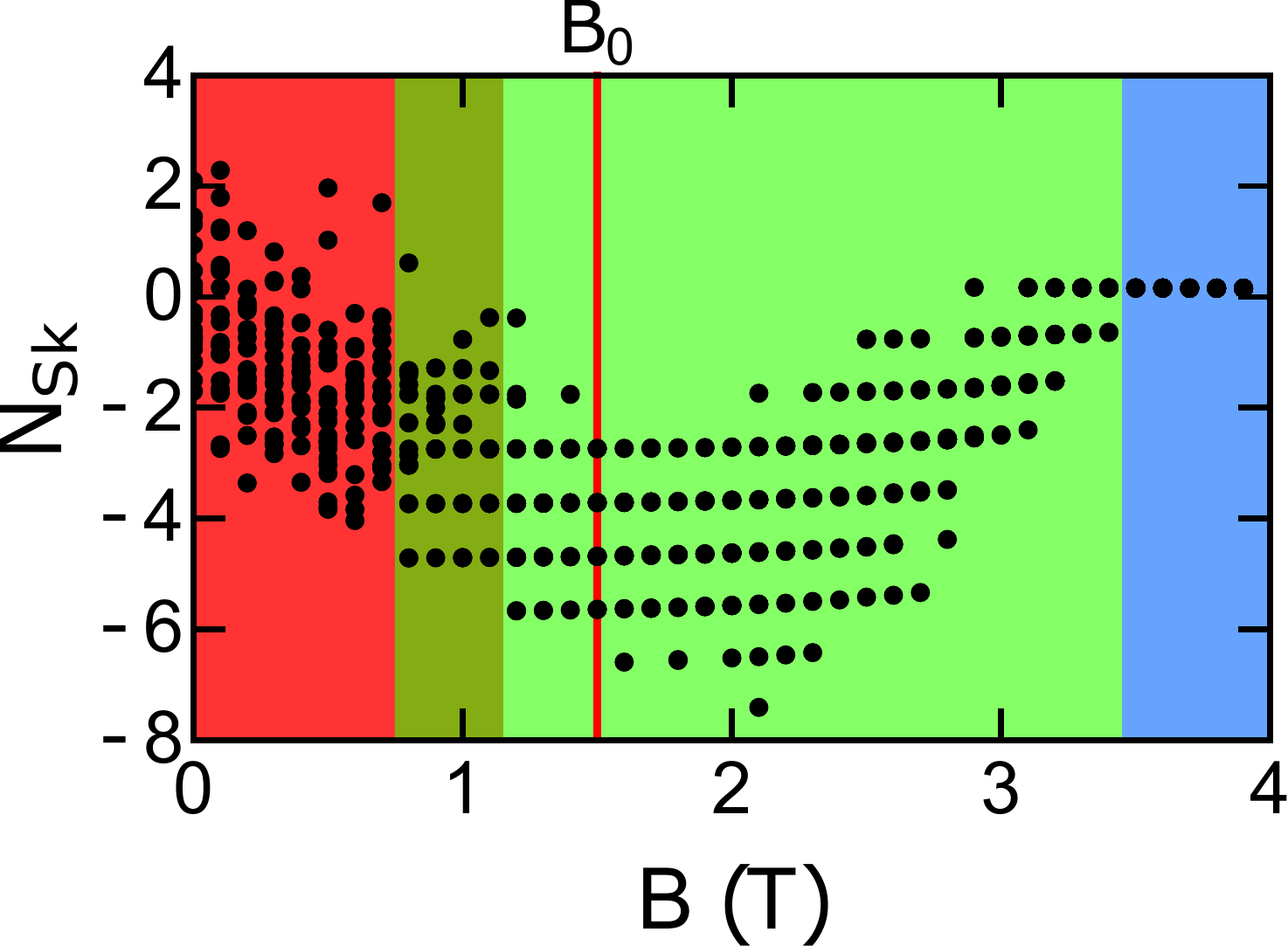}
		\caption{
			\textbf{Field dependence of the total topological charge}. 
			Red region: external fields below the strip-out instability leading to spin spiral segments; dark green: transition regime; bright green: ensemble of skyrmions; blue: completely field-polarized island, where skyrmions collapse due to the strong external field. Solid red line indicates the magnetic field $B=1.5\,$T used for subsequent calculations. For each step in the vertically applied field 20 randomly magnetized configurations of a  $21\,$nm diameter nanodisk were relaxed towards the nearest metastable state on the energy hypersurface at $T=0~$K. }\label{fig:1}
	\end{figure}
	We study metastable skyrmions and their motion and mobility at finite temperatures. At first the phase space has to be explored with respect to the external magnetic field and temperature. A detailed description of the phase diagram for extended systems of ultrathin biatomic layers of Pd/Fe on an Ir(111) substrate hosting magnetic skyrmions can be found, e.g., in Refs.~\cite{RoSi2016,bottcher2018b}. 
	Here, simulations are performed for the same system in the micromagnetic framework, by solving the Landau--Lifshitz--Gilbert equation (LLG)\cite{gilbert2004phenomenological}, as described in the Methods section.

	In Fig.~\ref{fig:1} possible equilibrium spin configurations are investigated  in a 21-nm-diameter nanodisk of $0.4\,$nm thickness at zero temperature. For each fixed external magnetic field ($B=B\ind{z}$), 20 different randomly generated initial states are considered and relaxed towards the closest local minimum of the total energy hypersurface. Subsequently, the total skyrmion number $N_\mathrm{Sk}$ of the relaxed states is calculated and plotted against the external magnetic field in Fig.~\ref{fig:1} to gain an overview of the possible metastable states.
	Below the strip-out instability field~\cite{leonov2016properties}, skyrmions as localized cylindrical spin structures are not stable and the configuration consists of spin spiral segments. For these the topological number is not a well-defined quantity, as spin spiral structures can extend over the whole island, and therefore the boundaries have a significant effect.  
	This is reflected by the continuous distribution of topological charge values found below $0.75$~T in the red regime in Fig.~\ref{fig:1}, whereas discrete skyrmion numbers correspond to discrete steps in the total topological charge.
	The upper boundary of this region is reasonably close to the strip-out field of $B=0.65~$T~\cite{leonov2016properties,rozsa2018localized} determined for extended systems.
	
	With increasing the external field, individual skyrmions may be stabilized in the system, where they will coexist with spin spiral segments. This corresponds to the darker green area in Fig.~\ref{fig:1} around $B\approx 1~$T, where besides the continuous distribution also discrete steps can be observed in the topological charge.
	
	In the following bright green region stronger magnetic fields lead to shrunken skyrmions, which in turn enable{\bcor s} the magnetic island to host a larger number of quasiparticles. In this regime the topological charge is always close to an integer, with the small deviation caused by the tilting of the spins at the edge of the sample.
	Reaching field values above $B\approx 3.5\,$T, the system becomes completely field-polarized and skyrmions are not stabilized anymore starting from a random configuration, even at zero temperature. Previous calculations~\cite{siemens2016minimal,leonov2016properties,rozsa2018localized} indicated that skyrmions on the lattice collapse at around $B=4.5~$T in the system.

\begin{figure}
		\includegraphics[width=.9\linewidth]{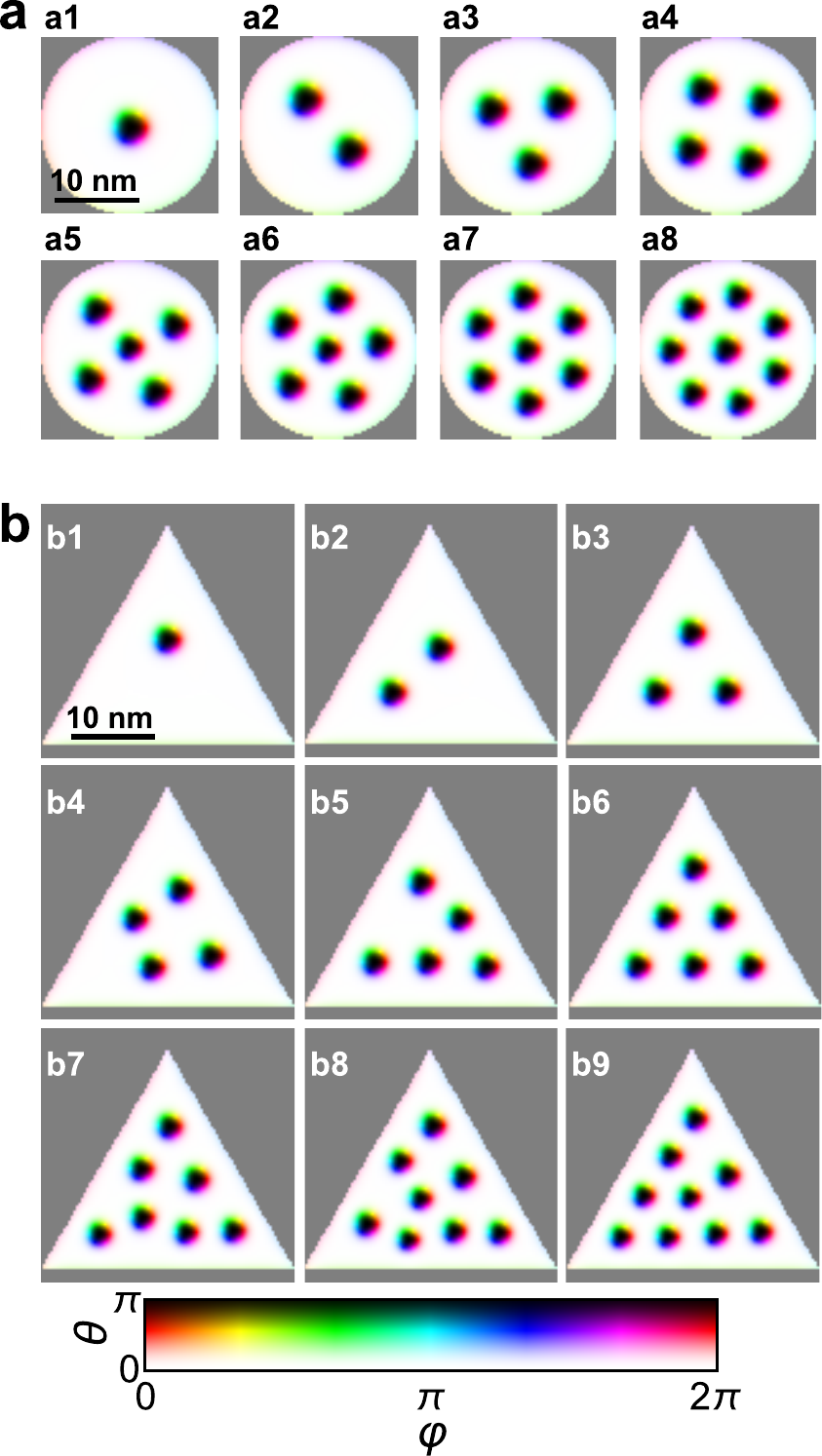}
		\caption{\textbf{Relaxed states for different initial skyrmion numbers.} Brightness corresponds to the out-of-plane component of the magnetization characterized by the polar angle $\theta$, whereas the color denotes the in-plane orientation described by the azimuthal angle $\varphi$, as shown in the color map. $B=1.5\,$T, $T=0\,$K. System size: \textbf{a} $21\times 21\times 0.4\,$nm$^3$; \textbf{b} $30\times 30\times 0.4\,$nm$^3$. The saturation magnetization is locally set to zero in the dark gray areas.
		}\label{fig:2}
	\end{figure}

	In order to prevent the appearance of spin spiral states and to be able to consider the skyrmions as well-defined quasiparticles, we choose $B=1.5\,$T for the following calculations, unless mentioned otherwise. 
	Figure ~\ref{fig:1} shows that several metastable configurations associated with different topological charges are accessible starting from randomly generated initial states. However, in this approach not all stable structures are covered, as the included ones are obtained from relaxing random initial configurations.  
	Different configurations may also be generated in a controlled way by adding skyrmions to the field-polarized state one-by-one, and relaxing the state at zero temperature. This is shown in Fig.~\ref{fig:2} for the disk-shaped and a triangular geometry for $B=1.5~$T.
		
\begin{figure}
		\includegraphics[width=.9\linewidth]{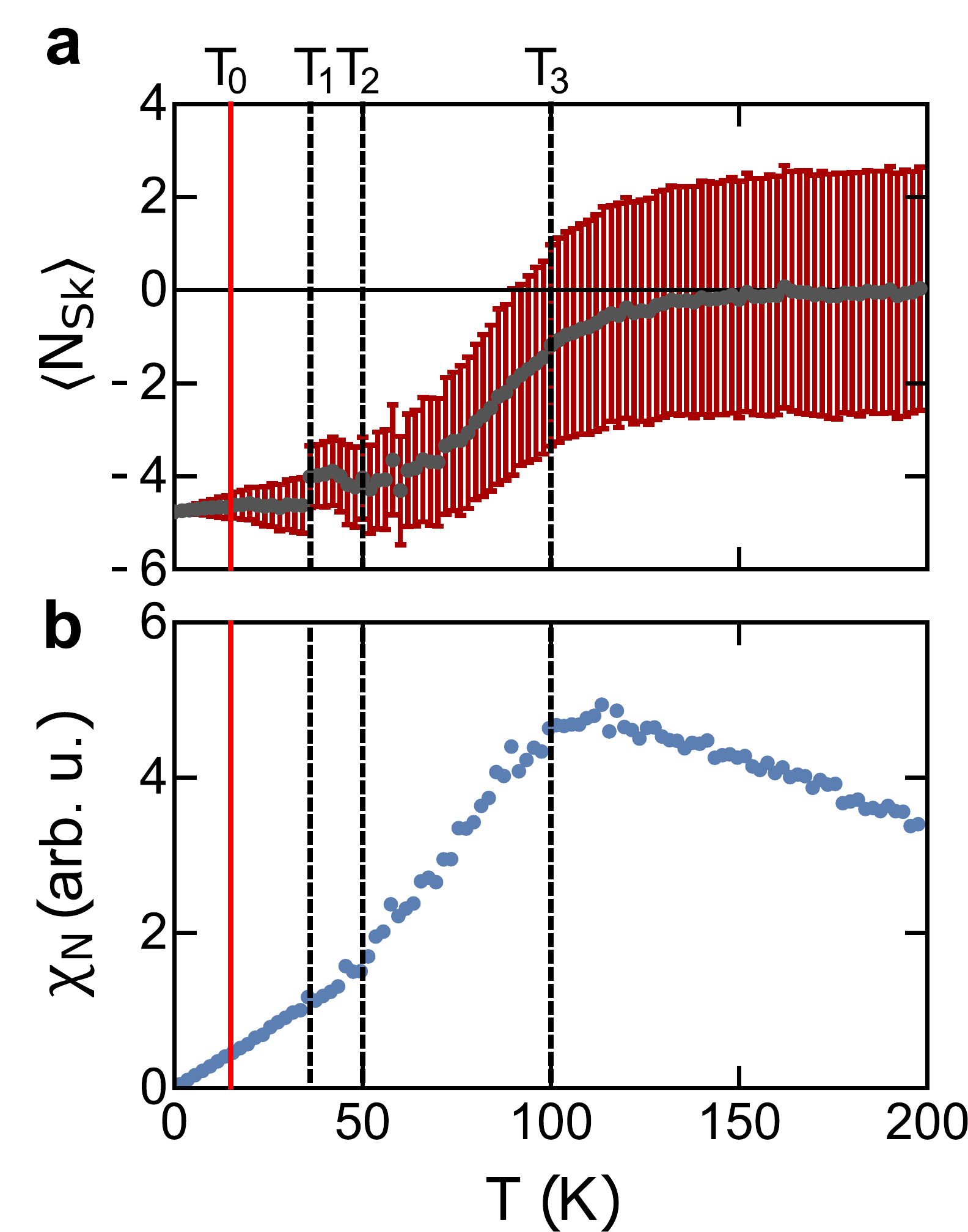}
		\caption{\textbf{Temperature dependence of the topological number.} Disk-shaped island  ($21\,\mathrm{nm}$ diameter) for $B=1.5\,$T. 
		\textbf{a} Total topological number $\langle N_\mathrm{Sk}\rangle$ averaged over $0.5\,$ns with error bars indicating the standard deviation. 
		\textbf{b} Topological number susceptibility  $\chi\ind{N}$ (Eq.~(\ref{eq:chiN})), with the dashed black lines indicating three characteristic temperatures discussed in the main text. Solid red line marks the temperature used for successive simulations. 
	 }\label{fig:3}
	\end{figure}

	On top of the transitions due to a variation of the magnetic field, the impact of finite temperature is also of crucial importance. 
	In Fig.~\ref{fig:3}a temperature effects on the topological charge are displayed. Starting from a relaxed configuration at $T=0~$K, the temperature is increased every $500\,$ps by $\Delta T=2\,$K and the topological charge is averaged over time at each temperature. 
	The results show a first discontinuity at $T_1\approx 40\,$K. 
	{\bcor Up to this temperature the thermal fluctuations are relatively weak, and the number of well-defined skyrmions inside the sample remains constant. The small standard deviation of the topological number in this regime can be attributed to the thermal motion of the spins at the edge.}
	However, the combination of thermal fluctuations and the repulsion between the skyrmions triggers the escape of a single skyrmion out of the sample at $T_1\approx 40\,$K. Hence, the total topological number is changed from $N_\mathrm{Sk}\approx-5$ to $N_\mathrm{Sk}\approx-4$, see Supplementary Video 1.
	This change is also shown in the plot of the \textit{topological number susceptibility} in Fig.~\ref{fig:3}b, defined as
	\begin{align}
		\chi_N=\frac{1}{T}\left(\langle N_\mathrm{Sk}^2\rangle - \langle N_\mathrm{Sk}\rangle^2 \right)~, \label{eq:chiN}
	\end{align}
	with $\langle N_\mathrm{Sk}\rangle =\frac{1}{t_1-t_0}\int_{t_0}^{t_1} N_\mathrm{Sk}(t)\dd t$. Only a minor deviation from the otherwise smooth function is visible. This means that the thermal fluctuations are still not strong enough to perturb the quasiparticles drastically.

	A first characteristic change in the slope of the topological susceptibility can be seen at $T_2\approx 50\mathrm{K}$. Comparing Figs.~\ref{fig:3}a and b it is obvious that the  fluctuations of topological number are increasing with temperature, and around this point the lifetime of skyrmions is reduced below the averaging window. The average topological charge continuously approaches zero above this temperature. In this disordered regime the lifetime of skyrmions is shorter than the observation time, as the fluctuations allow a collapse inside the sample. Hence, this temperature range is not favorable for information storage applications.

	 Finally, at about $T_3\approx 100\mathrm{K}$ the paramagnetic regime is entered, which is characterized by a completely disordered time-averaged magnetic configuration. Here the average topological charge is very close to zero. This behavior is also indicated by a change of sign in the first derivative  $\dd\chi_N(T)/\dd T|_{T=T\ind{3}}$. 
	 
	In the following, the external parameters are fixed to $T=T_0=15\,$K and $B=B_0=1.5\,$T, shown by the solid red lines in Figs.~\ref{fig:1} and \ref{fig:3}. This ensures that the thermal fluctuations do not lead to a collapse or escape for the skyrmions, only to a diffusive motion, and that the number of particle-like skyrmions will be a constant during the simulation time.
	
\begin{figure}
	\includegraphics[width=.9\linewidth]{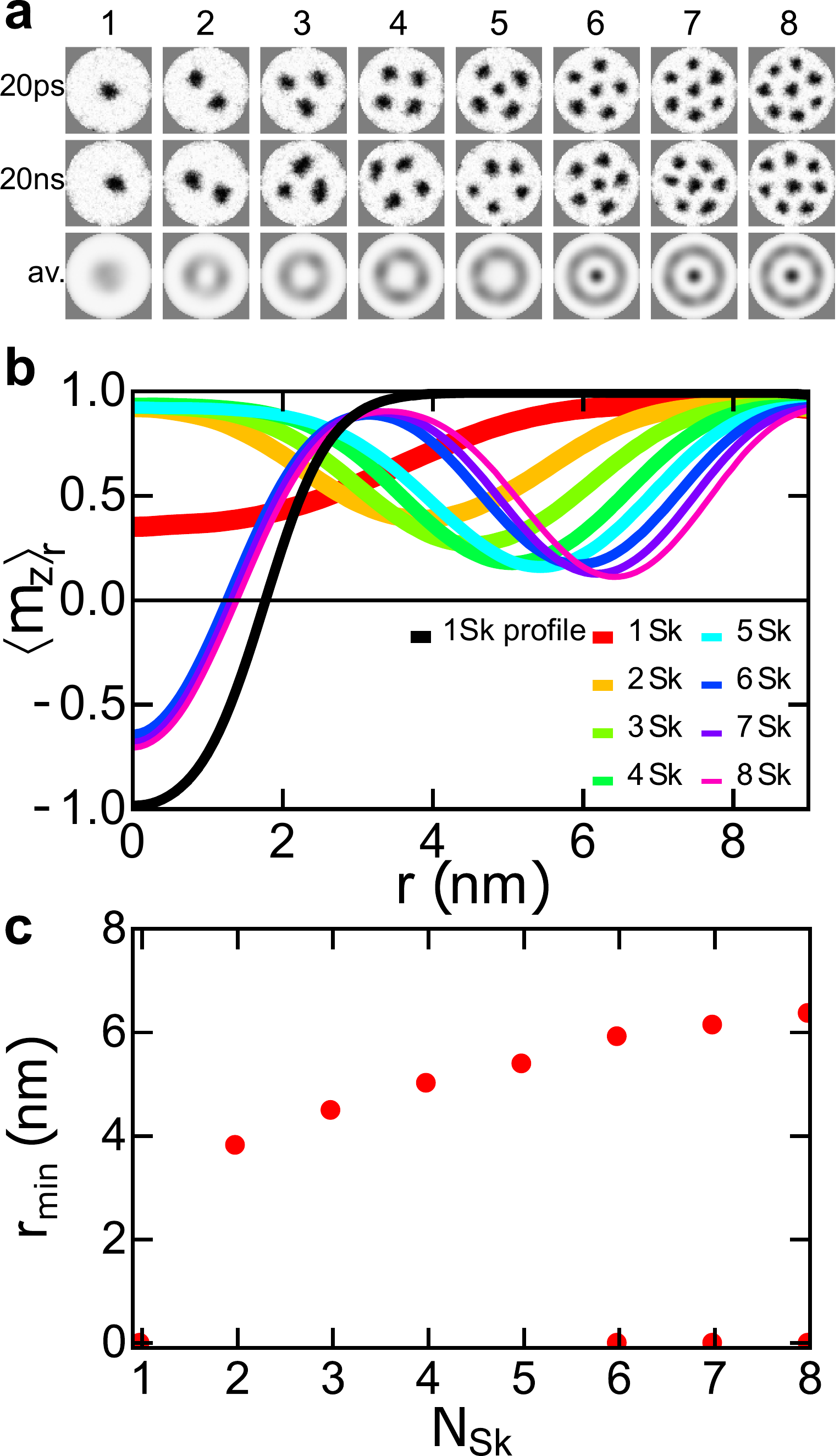}
	\caption{\textbf{Diffusive skyrmion motion on a circular island.}
		\textbf{a} $z$ component of the magnetization (white $+z$, black $-z$) for an ascending total skyrmion number. First row shows the configuration after $20\,$ps, the second row after $20\,$ns of elapsed time, and the third row represents the time average for the total simulation time over 1000 snapshots.
		\textbf{b} Time-averaged signal as a function of the distance from the center of the disk for different skyrmion numbers, averaged over the angular coordinate. Black line shows the static profile of a central single skyrmion.  
		\textbf{c} Radii $r\ind{min}$ corresponding to the local minima of the time-averaged signal in panel \textbf{b}, in dependence on the skyrmion number.   	
		Disk system ($21\times 21\times 0.4\,$nm$^3$), $T = 15~$K, $B=1.5~$T.  The saturation magnetization is locally set to zero in the dark gray areas. 
		} \label{fig:4}
\end{figure}

\begin{figure}
	\includegraphics[width=\linewidth]{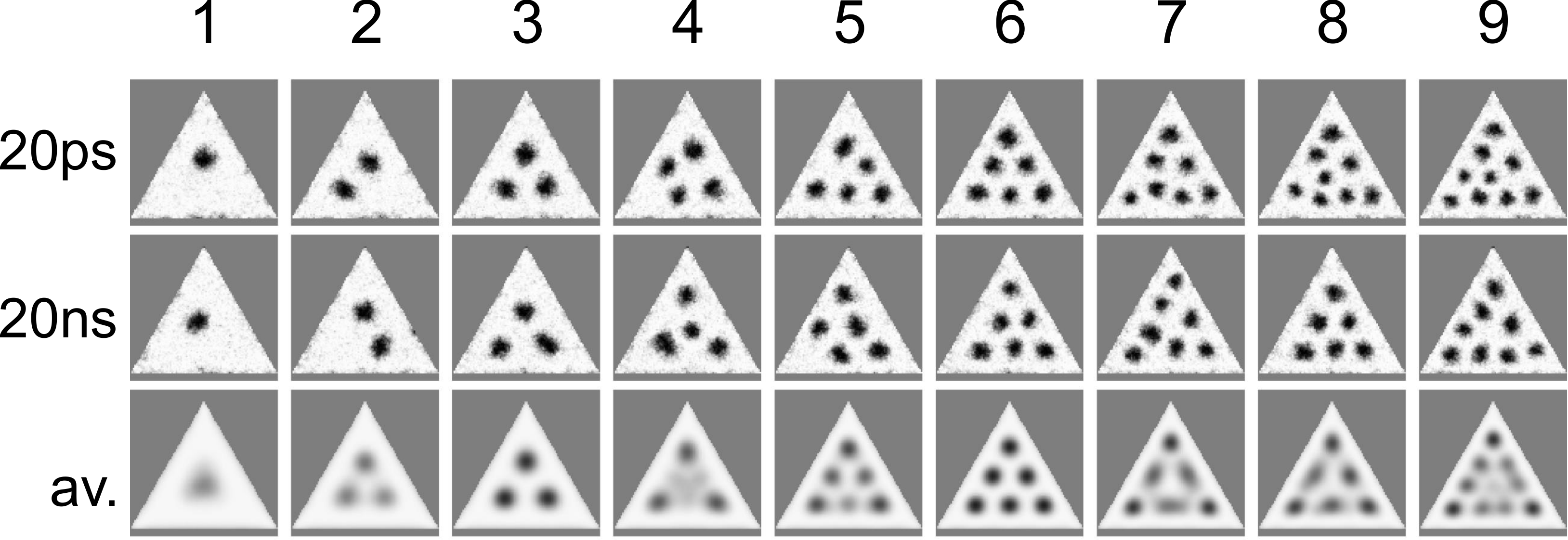}
	\caption{\textbf{Diffusive skyrmion motion on a triangular island.} $z$ component of the magnetization (white $+z$, black $-z$). First row shows the configuration after $20\,$ps, the second row after $20\,$ns of elapsed time, and the third row represents the time average for the total simulation time over 1000 snapshots. Triangular system ($30\times 30\times 0.4\,$nm$^3$), $T = 15~$K, $B=1.5~$T.  The saturation magnetization is locally set to zero in the dark gray areas.} \label{fig:5}
\end{figure}

\subparagraph{Skyrmion diffusion.}
	With these first results, giving information on the segment of the parameter space we deal with, we will focus on the influence of thermal fluctuations on the dynamics of the skyrmions.
	In Fig.~\ref{fig:4}a and \ref{fig:5} results for different initial numbers of skyrmions presented in Fig.~\ref{fig:2} are shown for the circular and triangular geometries, respectively. The first row shows the magnetic configuration after $20\,$ps of simulation time and the second row the final state, which means the magnetization state after our total simulation time of $20\,$ns. Only the out-of-plane $z$ component is shown in grayscale.  
	The magnetic configuration after only $20~$ps is qualitatively very similar to the final one, displaying slightly deformed skyrmion configurations due to the thermal fluctuations compared to the zero-temperature initial states in Fig.~\ref{fig:2}. {\bcor This short time scale on which the short-range Heisenberg and Dzyaloshinskii-Moriya interaction dominate the dynamics and cause shape deformations, is examined more extensively below.} 

	The third row corresponds to the time-averaged $z$ component, calculated over the simulation time divided into 1000 snapshots. In our understanding this result comes as close as possible to real-space scanning-probe measurements, due to the limited time resolution and finite measurement duration in such techniques. According to Ref.~\cite{leonov2016properties}, typical limits of the time resolution in SP-STM are $t_\mathrm{res}> 5~\mathrm{ms}$. 
	With the present simulation parameters we found that increasing the simulation time further does not lead to a significant change in the time-averaged images, indicating that similar observations may be expected on the much longer experimental time scales as well. 
	The observed two characteristic time scales can be attributed to different interaction types of clearly distinguishable energy scales: local deformations on the picosecond time scale of the magnetic texture are dominated by thermal excitations competing with the short-range symmetric and antisymmetric exchange interactions, {\bcor while the translational motion leading to} the complex pattern formation over tens of nanoseconds is caused by the repulsive interactions between pairs of skyrmions and between skyrmions and the boundaries.

	In case of the disk, it is extremely challenging to make a clear statement about the number of skyrmions in the system based on the time-averaged images only. 
	As the symmetry of the system is a cylindrical one, also the resulting picture is cylindrically symmetric, consisting of concentric bright and dark circles and rings. 
	The small contrast differences along the angular direction within a single diffusive ring-like area (cf. third row in Fig.~\ref{fig:4}a) arise because of the finite simulation time, but also as an artifact of the finite grid, leading to a deviation from the ideal radial symmetry. For all initial configurations the total skyrmion number is conserved. 
	Since the skyrmions repulse each other, the radius of the resulting gray ring increases for higher skyrmion densities, before one of them becomes localized quite strongly in the center. This behavior is analyzed in Fig.~\ref{fig:4}b, where the out-of-plane magnetization component depending on the distance from {\bcor the	} center of the disk is calculated by integration over the angular coordinate. For comparison, the static profile of a single skyrmion in the center of the disk is plotted as well (black curve). 
	The positions of the local minima of the integrated functions are shown in Fig.~\ref{fig:4}c.
	Without performing the averaging over time, the profile of the skyrmion rotates from $m\ind{z}=-1$ in the center towards $m\ind{z}=1$ in the field-polarized background. Due to the diffusive motion of the skyrmions, the time-averaged images show a smaller difference compared to the background magnetization at the approximate positions of the skyrmions, providing a measure for the localized nature of the quasiparticles in the sample. For an increasing number of skyrmions not only the radius of the resulting ring-shaped pattern increases, but so does the localization of the skyrmions.
	
	For the triangular system in Fig.~\ref{fig:5}, the lower symmetry of the geometry is reflected in the resulting time-averaged pictures. The symmetry of the spin configuration coincides with that of the sample for one, three and six skyrmions without thermal fluctuations, as shown in Fig.~\ref{fig:2}b. In the time-averaged images in the third row of  Fig.~\ref{fig:5}, this results in well-localized skyrmions with a strong dark contrast for these numbers of quasiparticles in the system.
	Interestingly, also the adjacent configurations show almost the same time-averaged pattern as the highly symmetric configuration, e.g. two or four skyrmions compared to the case of three skyrmions.
	For the case where one skyrmion is missing from the highly ordered configuration, the remaining skyrmions imitate the absent skyrmion and effectively show up as an additional ``phantom skyrmion''.
	In contrast, a surplus skyrmion smears out in the time-averaged picture, so that mostly the high-symmetry points remain observable, even though they can also be slightly broadened -- cf. the 6th and 7th columns in Fig.~\ref{fig:5}.
	For an even larger numbers of skyrmions, the repulsive interaction between them in combination with temperature fluctuations is strong enough for the skyrmions to escape at the boundary during the simulation time, as shown in the decreased number of quasiparticles in the second row compared to the first row in the 8th and 9th columns in Fig.~\ref{fig:5}.
	
	{\bcorr To identify the different time regimes more clearly, the convergence behavior of the time-averaged pictures is investigated. The time-integrated pictures are calculated up to each snapshot time and compared to the averaged image after $20~$nsvia a matching parameter, where a complete agreement is denoted by a value of $1$. {\bcor The mathematical definition of the matching parameter is given in the Methods.}
	The results are plotted in Fig.~\ref{fig:6}. Panels a and b show the convergence behavior for different skyrmion occupations for the disk and for the triangular geometry. For the circle there are no major differences in terms of the convergence speed for different skyrmion numbers. {\bcor Only the ensembles of three and seven skyrmions exhibit a slightly faster convergence, which is explicable by the resemblance of those states to the close-packed arrangement. As expected, the close-packed array possesses a lower mobility.} In the case of the triangle, for three and six skyrmions the value 1 is reached much faster than for the other configurations. This effect can again be explained by the strong localization of the skyrmions in these cases, preventing an exchange of their positions.
	A different trend is visible for the eight-skyrmion case (purple line in Fig.~\ref{fig:6}b), which shows a much slower convergence. This delayed behavior arises due to the non-conserved skyrmion number, which can be seen in Fig.~\ref{fig:5}. Consequently, after the escape of the skyrmions it takes more time until the averaged picture has adapted to this change. 
	
	For exploring the stochastic dynamics on the short time scale, {\bcor  which we suppose to be governed by the skyrmion deformations in contrast to the slow skyrmion displacement}, multiple simulations are performed for the same initial configuration, two skyrmions inside the disk relaxed at zero temperature, but with different pseudorandom number seeds. The simulations cover a time of $100$~ps each with 50 different initializations. After every $0.5$~ps a snapshot is taken and the positions of the two skyrmions extracted. On the scale of the simulation time no noteworthy displacement of each of the skyrmions takes place, however a deformation of the magnetic textures is visible. The mean skyrmion radius relative to the zero temperature radius {\bcor calculated as described in the Methods,} is averaged over the different simulations, shown in Fig.~\ref{fig:7}. Here, two features are remarkable. 
	Firstly, the radius increases, approaching an enhancement of about $10$~\% after $100$~ps. 
	Secondly, the radius is oscillating with a high frequency of approximately $60$~GHz, estimated from the average time between the peaks. This oscillation can be identified with internal skyrmion modes, like the so-called breathing modes {\bcor(cf. Ref.~\cite{rozsa2018localized}: at $T=0\,$K, $B=1.5\,$T and the triangular atomic lattice the breathing mode frequency is $f=70.88\,$GHz, which is expected to decrease at higher temperature)}.
	These findings support the result that in confined geometries, different time scales are important for the skyrmion dynamics. On the short time scale the internal modes and deformations of the skyrmions are dominant, whereas on the longer time scale the interaction leads to a slowly converging pattern formation.
	\begin{figure}
	\includegraphics[width=.9\linewidth]{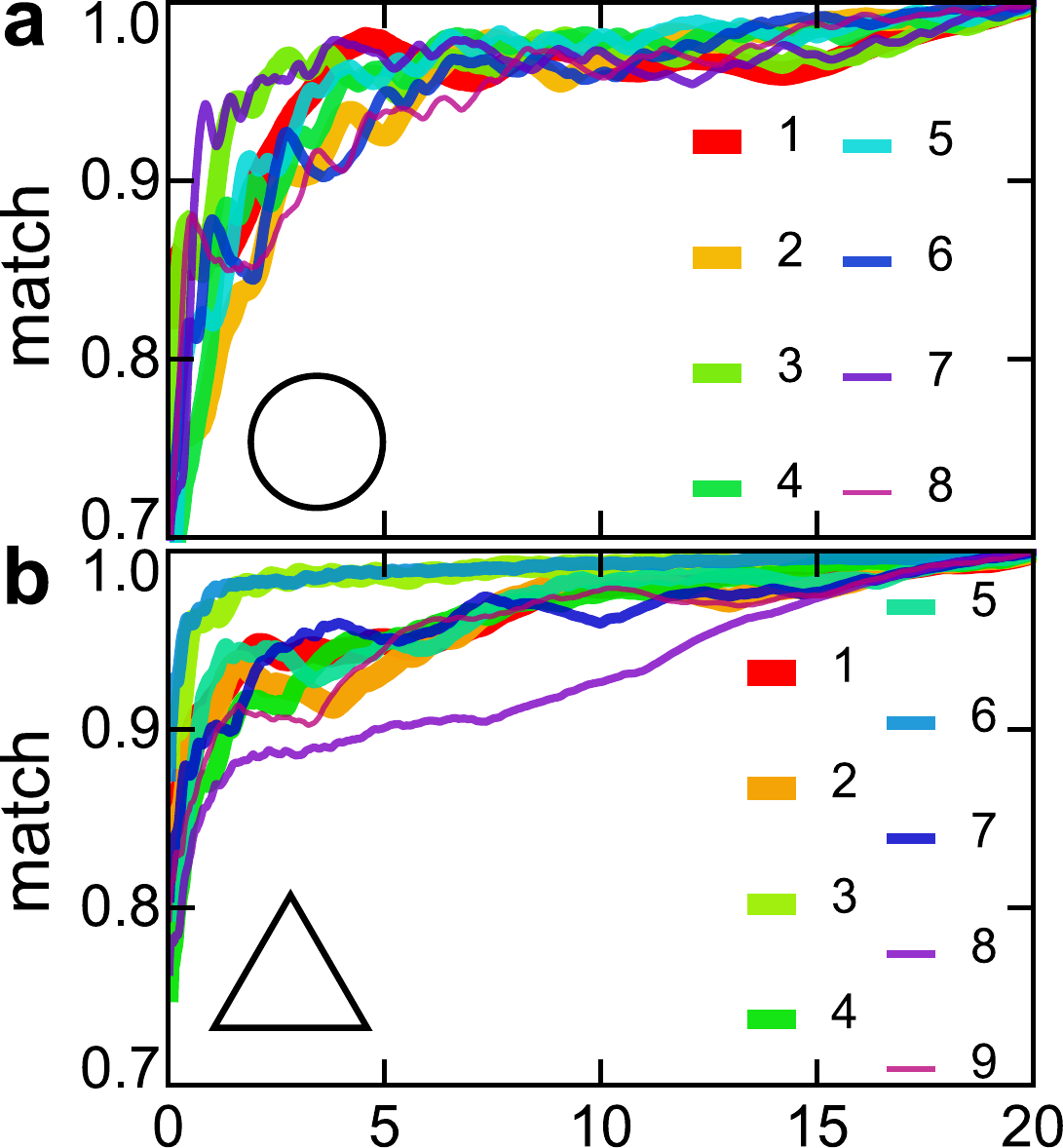}
		\caption{
			{\bcorr\textbf{Convergence behavior of the time-integrated magnetization for different skyrmion numbers.}	
			The match between the final time-averaged pattern after $20\,$ns and the time average taken until time $t$ is defined such that a value of $1$ corresponds to perfect agreement, see the Methods for details. \textbf{a} and \textbf{b} show the convergence of the matching parameter for the disk and triangular setup, respectively, for different skyrmion numbers.
			}
		}\label{fig:6}
	\end{figure}

	\begin{figure}
	\includegraphics[width=\linewidth]{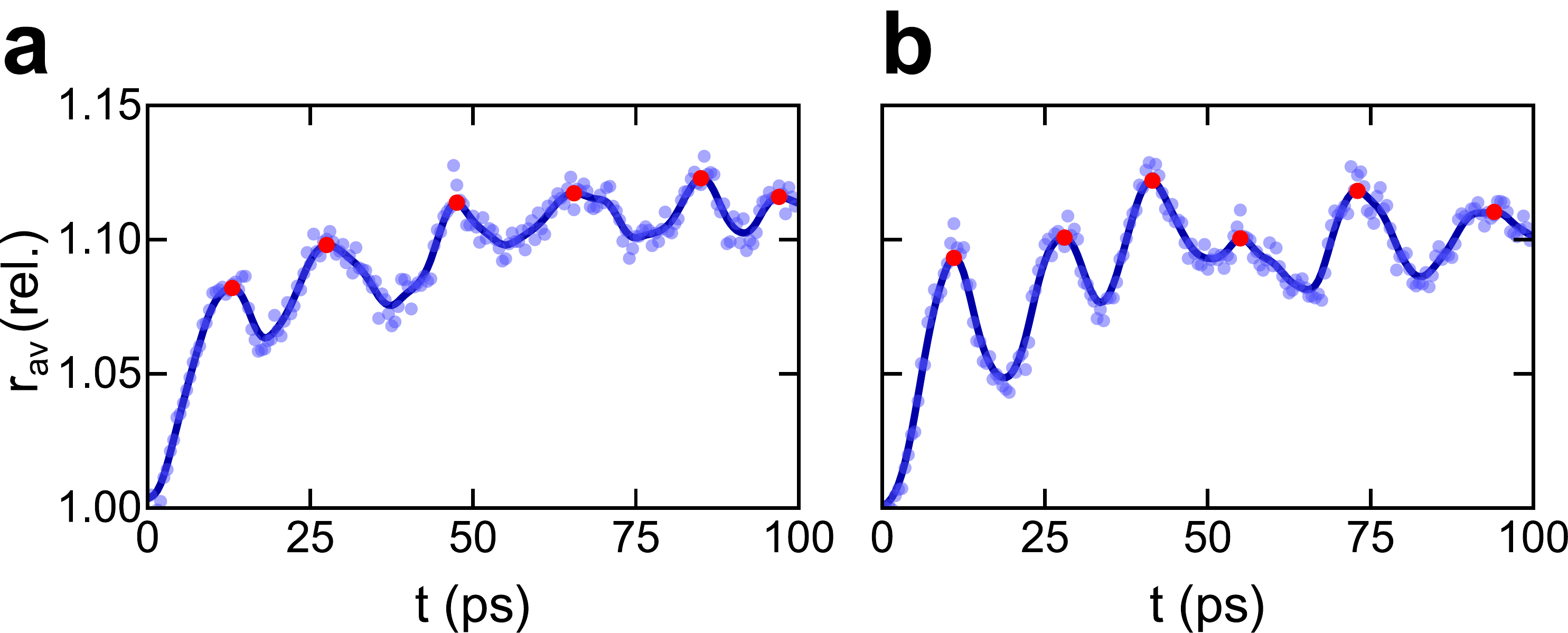}
	\caption{
		{\bcorr \textbf{Fast deformation dynamics of two skyrmions on a disk-shaped island.}	
		Two skyrmions on a disk $(21\times21\times0.4)\,$nm$^3$ are initialized from a relaxed zero-temperature configuration. Afterward, 50 different seeds for the pseudorandom number generator are used to simulate the skyrmion dynamics over $100\,$ps, saved every $0.5\,$ps. The results are analyzed by calculating the centers of the skyrmions, followed by the computation of the average skyrmion radius (see Methods) for each of them as shown in \textbf{a} and \textbf{b}. Averaging over the different simulations smears out most of the pure stochastic dynamics, yielding an oscillating pattern superposed with a general trend of an increased skyrmion radius compared to the zero-temperature configuration. Bright blue points show data averaged over different simulations, the solid blue line corresponds to data smoothened by a Gaussian filter in order to calculate the peak positions (red dots).}
	}\label{fig:7}
\end{figure}

}
	
\begin{figure}
		\includegraphics[width=.9\linewidth]{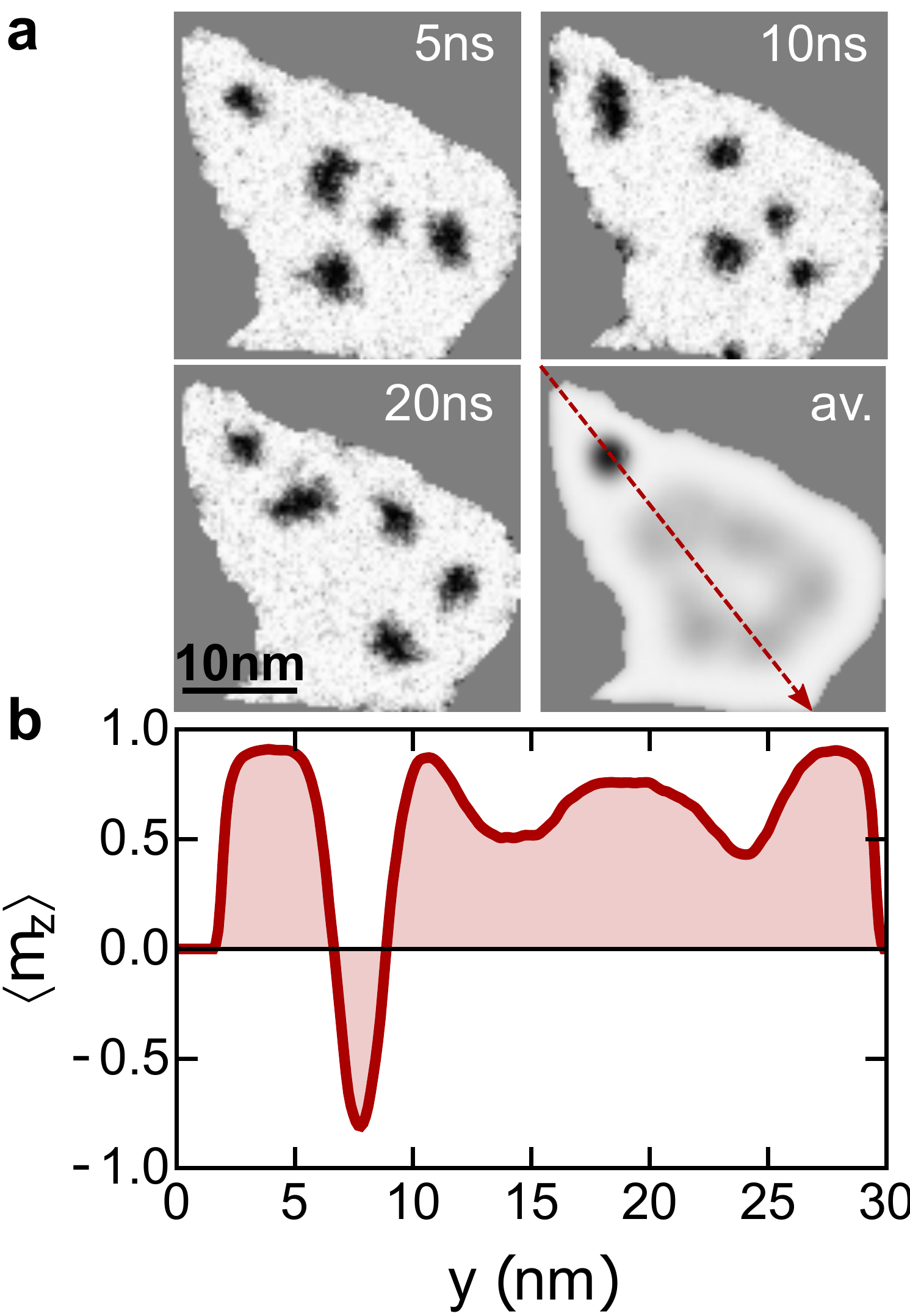}
		\caption{
		\textbf{Diffusive skyrmion motion on an asymmetric nanoisland including a defect.}	
		$z$ component of the magnetization (white $+z$, black $-z$) is imaged. In the top-left corner a defect is modeled by freezing the magnetization direction in a single simulation cell along the $+x$ direction. \textbf{a} Snapshots of the magnetization at different times, as well as the resulting time-averaged picture (\textbf{av.}).
		\textbf{b} Time-averaged signal along the dashed arrow shown in the lower right part of panel \textbf{a}. 
Simulation box: $30\times 30\times 0.4\,$nm$^3$, $T=25~$K, $B=1.5~$T. The saturation magnetization is locally set to zero in the dark gray areas.
		}\label{fig:8}
	\end{figure}

	All of the discussed examples showcase the strong influence of the interaction between skyrmions as well as of the geometrical aspects on their mobility, that can vary between delocalization and a strong localization, e.g., in the center of the disk in Fig.~\ref{fig:4}a.
	This feature of variable mobility, depending on the geometry and packing density of skyrmions, is of general nature. In real experimental systems further contributions may affect the mobility, and ultimately the measured position of the skyrmion as well. As an example of such a feature, defects in the grown thin film nanoislands will be discussed.   
	In Fig.~\ref{fig:8}, a nanoisland geometry inspired by the experimental results in Refs.~\cite{romming2013writing,RoKu2015} is considered. Additionally to the previous simulation parameters, a magnetic defect is inserted in the top-left corner of the island, treated as a simulation cell with the magnetization frozen along the in-plane $+x$ direction. The initial configuration including 5 skyrmions is evolved in time over a span of $20~$ns, as shown in the snapshots in panel a. During the time evolution one skyrmion is pinned at the defect, whereas the others are moving rather freely. The resulting time average of the images reflects this behavior, as only a single spot is visible with a well-defined position in the upper left corner, and the rest of the quasiparticle features form a blurred trace mostly following the geometry of the island. 
	Additionally, the profile of the out-of-plane spin component is shown in Fig.~\ref{fig:8}b along the dashed arrow in the lower right panel of Fig.~\ref{fig:8}a. In this representation the discrepancy between the measured profiles of the different skyrmions is even more pronounced.
	Where the localized skyrmion appears as a sharp dip almost reaching $m_{z}=-1$, the superposition of the other four skyrmions leads to broader and shallower features, and therefore no clear indication of their positions. 

\subparagraph{Quasiparticle model.}
	 The localization of the quasiparticles discussed above may also be interpreted as the probability density of finding a skyrmion at a selected position over the complete simulation time. Such a probabilistic interpretation is capable of describing the different contrast levels observed for nominally equivalent skyrmions, and it proves to be valuable in the interpretation of scanning-probe experimental results where the characteristic measurement times are significantly longer than the time scale of the diffusive motion. In this section, a simplified quasiparticle model is introduced, motivated by the time-averaged data of the micromagnetic simulations, offering a time-efficient opportunity in predicting the complex pattern formation of the skyrmion probability distribution discussed above.
	 
 	For the purpose of using a quasiparticle approach, the repulsive potential between pairs of skyrmions and between the skyrmion and the boundary needs to be quantified.
	Due to the Neumann boundary condition~\cite{VaLe2014} at the edge of nanoislands, the DMI leads to a noncollinear spin texture, similar to a skyrmion~\cite{rohart2013skyrmion}. Therefore, the interaction mechanism is the same between the skyrmions and between a skyrmion and the boundary, as has been studied in previous publications~\cite{LiRe2013}.
	{\bcorr For details on the calculation and the explicit functions of the potentials see the Methods section. }
		
	Starting from the modeled skyrmion-skyrmion and skyrmion-boundary potentials quasiparticle simulations are executed, in order to compare the calculated probability densities with the time-averaged configurations obtained from full-fledged micromagnetic simulations. 
	{\bcorr The main assumption for this endeavor is that neither the thermal fluctuations lead to the collapse, creation or escape of skyrmions, nor does a high skyrmion density {\bcor lead to a merging of the skyrmions} on the island as discussed in Ref.~\cite{siemens2016minimal}.}
	A triangular setup containing two quasiparticles serves as the model system. The side of the triangle is chosen to be $30\,$nm, the same as in the case of the micromagnetic calculations shown in Figs.~\ref{fig:2}b and \ref{fig:5}.
	The motion of the skyrmions inside the potential is calculated by modeling them as interacting random walkers, using an implementation of a Markov-chain Monte Carlo algorithm with Metropolis transition probabilities to study their diffusion, described in detail in the Methods section. This method offers a fast way of obtaining the cumulative probability density function of the positions of the two particles, which in turn can be compared to the time-averaged images in Figs. \ref{fig:4} and \ref{fig:5} obtained from micromagnetic simulations.
	{\bcorr In the following, a quantitative comparison between full micromagnetic simulations and Monte Carlo calculations will be presented to review the validity of the simplified quasiparticle approach. For this endeavor, we tracked the positions of two skyrmions in the triangular system from the micromagnetic simulation snapshots after every $20$~ps and calculated the probability density distribution of the center coordinates. The resulting distribution along with the Monte Carlo result after $10^5$ Monte Carlo steps is shown in Fig.~\ref{fig:9}.
	The largest difference between the two results arises because of the substantially {\bcor less data} for the micromagnetic calculation. As the histogram is calculated with only $10^3~$snapshots compared to $10^5$ Monte Carlo steps, this behavior is expected. Nevertheless, the characteristic features are similar. In both results the three peak densities are not radially symmetric, but are resembling the triangular boundary shape. 
	Hence this deformation is not an artifact of the Monte Carlo simulation. 
	Additionally, the average distance between the density maxima may serve as a good quantity for comparing both methods. 
	As Fig.~\ref{fig:9}b lacks the resolution for calculating the peak positions reliably, we start instead from the time-integrated magnetization pictures in Fig.~\ref{fig:5}. Based on the averaged result for two skyrmions, we extract again the peak positions and calculate subsequently the mean distance. 
	The positions for the Monte Carlo calculations are at $(11.0,6.5)\,$nm, $(15.0,13.5)\,$nm, $(19.0,6.5)\,$nm, the micromagnetic simulations deliver $(11.1,6.6)\,$nm, $(15.0,13.5)\,$nm, $(19.0,6.5)\,$nm. This leads to an average distance of $8.0\,$nm for both the Monte Carlo calculations and the micromagnetic simulations, {\bcor supporting} the validity of the simplified quasiparticle approach in the investigated parameter space. 
	
	The model is capable of predicting the skyrmion distribution in the long time limit and {\bcor thereby serves} as a method for computationally efficient calculations for larger or more complex systems.
	\begin{figure}
		\includegraphics[width=\linewidth]{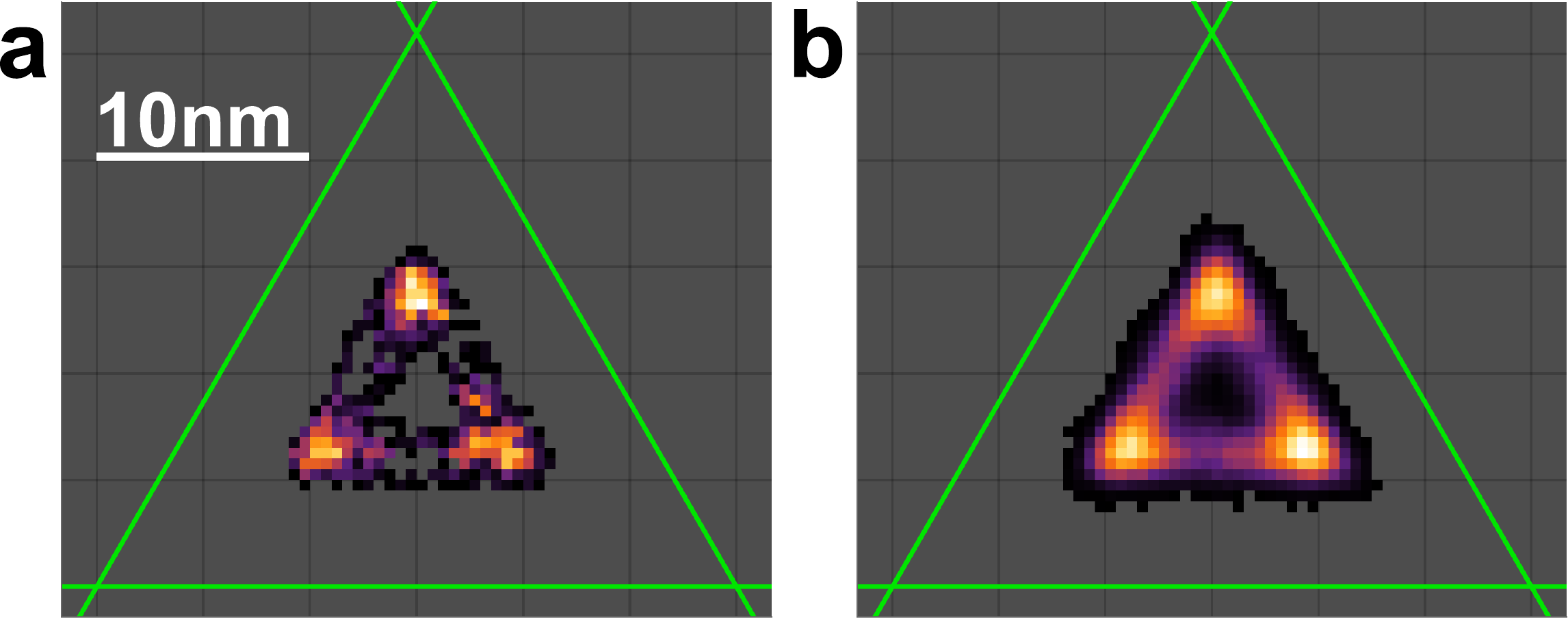}
		\caption{
			{\bcorr\textbf{Comparison of the Monte Carlo and micromagnetic simulations. }
				The probability density function for two skyrmions inside a triangular island as obtained from {\bcor \textbf{a}:} micromagnetic simulations, and {\bcor\textbf{b}:} from calculations following the Monte Carlo algorithm. For the micromagnetic distribution, the coordinates of the skyrmions are extracted from the snapshots taken every $20~$ps over $20~$ns of simulation time at $T=15\,$K.  
				Panel \textbf{b} is generated over $10^5$ Monte Carlo steps.}
		}\label{fig:9}
	\end{figure}
	}
	%
	As an example, the probability density functions of the Monte Carlo simulations are shown in Fig.~\ref{fig:10} after {\bcorr $10^5$} Monte Carlo steps for different temperatures and sample sizes. 
	Due to the repulsion between the two skyrmions, equilibrium positions are found close to the vertices of the triangle, with energy barriers between these local energy minima.
	At low temperature ($T=1\,$K) and a small island sizes ($30\,$nm edge length) in the upper left panel in Fig.~\ref{fig:10}, no transitions between the minima occur during the duration of the simulation, and two skyrmions may be observed in the obtained probability densities.
	As the temperature is increased to $T=15\,$K or $T=30\,$K in the first row of Fig.~\ref{fig:10}, the transition rate between the preferential positions is enhanced, and an additional ``phantom skyrmion'' appears in the probability density function, in remarkable agreement with  the micromagnetic results as analyzed before.	
	{\bcor	By enlarging the size of the island (second and third rows in Fig.~\ref{fig:10} for $45\,$nm and $60\,$nm edge length), the energy surface becomes flatter, which in turn leads to higher transition probabilities between the energy minima at the same temperature.
	Therefore, raising the temperature or increasing the size of the system have similar effects on the resulting probability density functions for the skyrmions. Consequently, large island sizes or high temperatures result in completely delocalized skyrmions as indicated in the bottom right panel in Fig.~\ref{fig:10}.  
	}
	\begin{figure}
		\includegraphics[width=\linewidth]{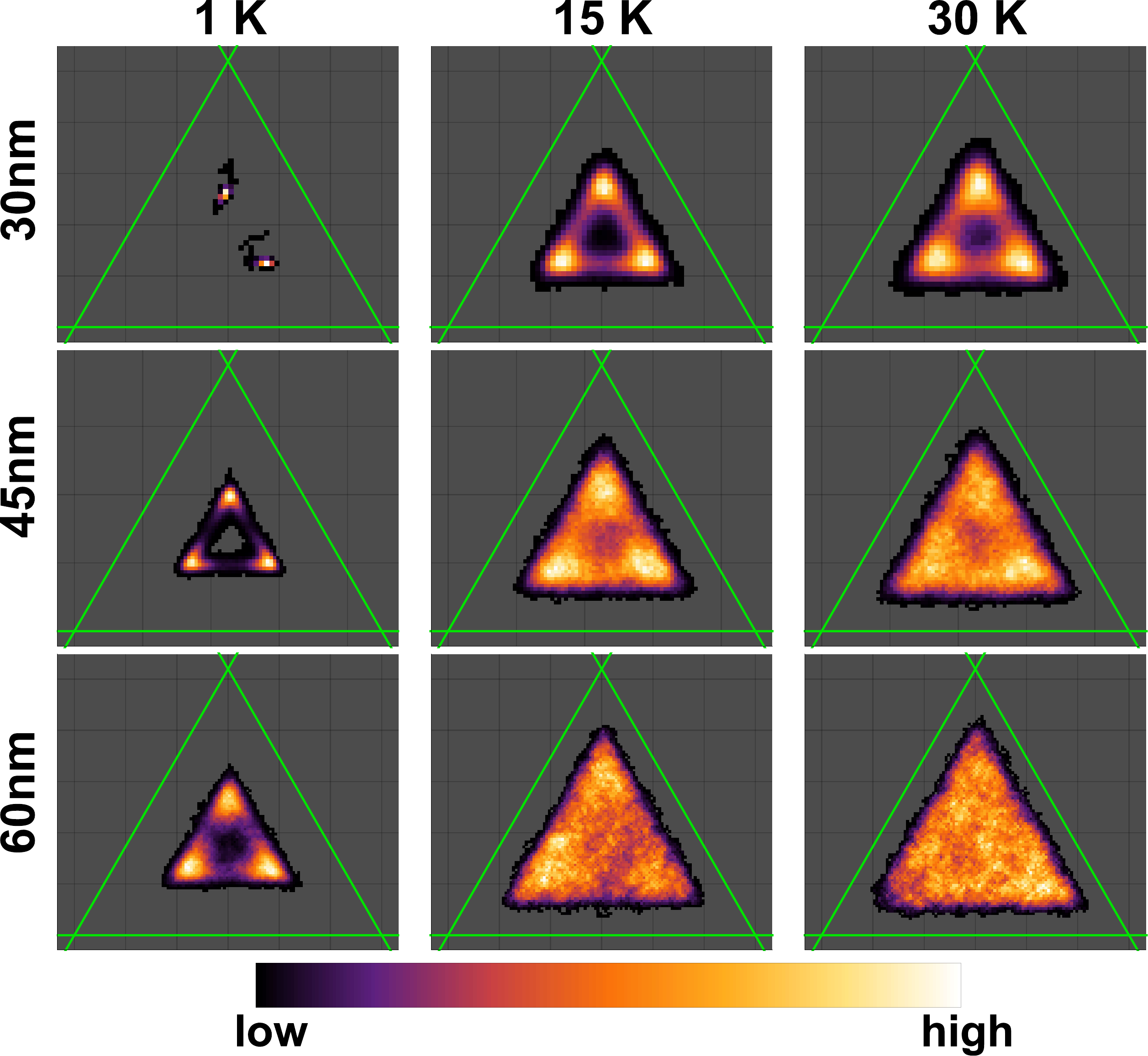}
			\caption{
				\textbf{Probability density function for two skyrmions inside a triangular model potential calculated following a Monte Carlo algorithm.}
				{\bcorr The temperatures are $T=1~$K, $T=15\,$K and $T=30\,$K, the edge lengths of the triangle are $30\,$nm, $45\,$nm and $60\,$nm.  Probability distributions for the centers of the two skyrmions are calculated over $10^5$ Monte Carlo steps.}
		}\label{fig:10}
	\end{figure}
%

\section*{Discussion}
	In this paper we studied the temperature-driven diffusive motion of ensembles of magnetic skyrmions in finite magnetic islands. 
	Based on experimentally determined system parameters for ultrahin Pd/Fe bilayers on Ir(111), we performed full-fledged micromagnetic simulations for various nanoisland geometries. For moderate temperatures and external magnetic fields, magnetic skyrmions with a long lifetime are present in the system, so that the topological charge is conserved. 
	Our study of this model system showed two different time scales on which the stochastic dynamics of skyrmions takes place.
	On the short time scale ($t \sim 1\,\mathrm{ps}$), the fluctuating dynamics governed by the symmetric and antisymmetric exchange interaction is most prominent, which essentially leads to local deformations, {\bcorr exciting internal modes of the skyrmions}. In the long-time limit ($t \gtrsim 10\,\mathrm{ns}$), interactions between pairs of skyrmions, as well as between the quasiparticles and the boundaries, are dominant. These thermally activated mechanisms lead to complicated time-averaged pictures of the skyrmion distribution, where the number of quasiparticles in the system is not immediately apparent. 
	These results can be compared directly to images obtained from time-integrating measurement techniques such as SP-STM or MFM, the time-resolution of which ($\sim 5~\mathrm{ms}$~\cite{leonov2016properties}) is typically much longer than the timescale of the inherent dynamics of skyrmions. 
	Our findings open an alternative way compared to conventional interpretations of experimental results obtained with scanning probe methods on {\bcorr mobile} arrays of skyrmions, by differentiating between the various time scales of magnetization dynamics. 
	
	Furthermore, we proposed and developed a Monte Carlo quasiparticle model, which can be of great use to efficiently calculate the stochastic motion of larger systems of skyrmions in arbitrarily shaped geometries. 
	According to the results, the time-averaged images of the micromagnetic simulations can be qualitatively well reproduced by the probability density distributions obtained from the simple quasiparticle model, as long as it is possible to treat the skyrmions as stable objects with lifetimes significantly longer than the simulation time at the given temperature. 
	This method can be used to describe larger systems containing more skyrmions at various temperatures in a computationally efficient way, including model applications in storage technology like racetrack memory devices. Here, not only the stability of the information bit is important, but also its addressability, which is immediately affected by the diffusive motion of the skyrmions.

\section*{Methods}
\subparagraph{Micromagnetic simulations.}
	Finite-temperature micromagnetic calculations, using the open-source, GPU-accelerated software package \texttt{mumax3}\cite{VaLe2014}, were performed to solve the LLG equation,
	\begin{align}
	\dot{\vec{m}}_i(t)=-\frac{\gamma}{1+\alpha^2}\left[\vec{m}_i\times\vec{B}_{i}^\mathrm{eff}+\alpha\vec{m}_i\times\left(\vec{m}_i\times\vec{B}_{i}^\mathrm{eff}\right) \right]~,
	\end{align}
for every simulation cell $\vec{m}_i$ of the discretized magnetization vector field. 
	Here, $\gamma_0=1.76\times 10^{11}($T$^{-1}$s$^{-1})$ is the gyromagnetic ratio of an electron and $\alpha$ is the Gilbert damping parameter. The time- and space-dependent effective magnetic field,
	\begin{align}
		\vec{B}_i^{\mathrm{eff}}(t)=\vec{B}_i^\mathrm{ext}+\vec{B}_i^\mathrm{exch}+\vec{B}_i^\mathrm{d}+\vec{B}_i^\mathrm{a}+\vec{B}_i^\mathrm{dmi}+\vec{B}^\mathrm{th}_i\left(t\right)~, \label{eq:heff}
	\end{align}
	is composed of the external field $\vec{B}_i^\mathrm{ext}$; exchange interaction field  $\vec{B}_i^\mathrm{exch}=2A\ind{exch}/M\ind{sat}\Delta\vec{m}_i$, with $A\ind{exch}$ the exchange stiffness and $M\ind{sat}$ the saturation magnetization; demagnetizing field $\vec{B}_i^\mathrm{d}=M\ind{sat}\hat{\vec{K}}_{ij}*\vec{m}_j$, where details on the calculation of the demagnetizing kernel $\hat{\vec{K}}$ can be found in Ref.~\cite{VaLe2014}; uniaxial magnetocrystalline anisotropy field $\vec{B}_i^\mathrm{a}=2K\ind{u}/M\ind{sat} m_z\vec{e}_z$, with $K\ind{u}$ the anisotropy constant; and the field generated by the Dzyaloshinskii-Moriya interaction $\vec{B}_i^\mathrm{dmi}=2D/M\ind
	{sat}(\partial m_z/\partial x,\partial m_z/\partial y,-\partial m_x/\partial x-\partial m_y/\partial y)^T$, with $D$ the strength of the interfacial Dzyaloshinskii-Moriya interaction. The simulation parameters were determined experimentally in Ref.~\cite{RoKu2015} based on the field dependence of the skyrmion profile:
$M\ind{sat}=1.1\,$MA/m, 
$D=3.9\,$mJ/m$^2$, 
$A\ind{exch}=2\,$pJ/m,
$K\ind{u}=2.5\,$MJ/m$^3$
and $\alpha = 0.05$.

Additionally, an effective thermal field is included in Eq.~(\ref{eq:heff}) as
	\begin{align}
		\vec{B}_i^\mathrm{th}=\vec{\eta}\sqrt{\frac{2\alpha k\ind{B}T}{M\ind{sat}\gamma_0\Delta V \Delta t}}\, ,
	\end{align}
	where $\vec{\eta}$ is a random vector generated according to a standard normal distribution independently for each simulation cell and time step. $k\ind{B}$ is Boltzmann's constant, $T$ is the temperature, $\Delta V$ is the size of the simulation cell and $\Delta t$ is the time step of simulation. 

	During the simulations, the cell size was set to $\Delta V=0.3\times0.3\times0.4\,$nm$^3$. The linear size of the cell is comparable to the lattice constant $a=0.271\,$nm of the Pd/Fe bilayer on Ir(111). This means that the micromagnetic simulations performed here should closely resemble the results of atomistic simulations where the skyrmions collapse when their size becomes comparable to the lattice constant~\cite{siemens2016minimal,leonov2016properties}, and where the use of temperature-independent model parameters is justified.

	The total skyrmion number $N\ind{Sk}$ in the simulations was calculated via
	\begin{align}
		N_\mathrm{Sk}= \frac{1}{4\pi}\int \vec{m}\cdot\left(\frac{\partial \vec{m}}{\partial x}\times\frac{\partial\vec{m}}{\partial y}\right)\dd x \dd y~.
	\end{align} 
{\bcor\subparagraph{Processing of the micromagnetic results.}
	For the results shown in Fig. 6, the z component of the magnetization averaged up to time $t$ is stored in grayscale pictures as a matrix, $A_{ij}(t)=(m_{z}(x_{i},y_{j},t)+1)/2$, where $x_{i}$ and $y_{j}$ denote the position of the micromagnetic simulation cell. The distance induced by the Frobenius norm between the image averaged up to time $t$, $A_{ij}(t)$, and the image averaged over the whole simulation length $\tau=20\,$ns, $A_{ij}(\tau)$ is divided by the square root of the number of matrix elements and subtracted from 1, yielding the matching parameter
	\begin{align}
		M(t) = 1 -  \sqrt{\sum_{i=1}^{N_x}\sum_{j=1}^{N_y}\frac{(A_{ij}(t)-A_{ij}(\tau))^2}{N_x N_y}}\, .
	\end{align}
	This procedure gives a matching parameter of $1$ for a perfect agreement of the compared pictures. 

	 In order to determine the skyrmion radius as a function of time in Fig. 7, firstly the contour lines where the z component of the magnetization is zero are calculated for the skyrmion at each time step. The contour lines are discretized on N points in space, $\vec{L}_{i}(t)=(x_{i}(t),y_{i}(t)), i\in(1,...,N)$. Subsequently, the center of the skyrmion is calculated via
	\begin{align}
		\vec{c}(t)=
		\frac{1}{N}\sum_{i=1}^{N}\vec{L}_i(t)~,
	\end{align}
	which in turn is used to obtain the average skyrmion radius
	\begin{align}
		\bar{r}(t) = \frac{1}{N}\sum_{i=1}^{N}||\vec{L}_i(t)-\vec{c}(t)||~.
	\end{align}
	Finally, the radius is normalized with respect to the zero temperature radius $r_0$ obtained from the relaxed initial configuration of choice.
}
\subparagraph{Calculation of the skyrmion-skyrmion and skyrmion-boundary interaction potentials.}
 	For the calculation of the potential, the total energy of a stripe-shaped model system ($75\times 30\times 0.4\,$nm$^3$) was investigated with micromagnetic simulations for two different cases. 
	In the first case, Neumann boundary conditions were used along the long side, and a homogeneous magnetization was relaxed leading to a canted rim. Afterwards a skyrmion was added to the system at a certain distance from the boundary.  
	Because of the repulsive nature of the interaction, a standard relaxation of the skyrmion position was not suitable for determining the distance dependence. Instead, we fixed the central magnetic moments in the skyrmion (in a circular region of $0.9$~nm diameter) at a specific position and relaxed the magnetic texture in the other cells. Analogously, in the second case the skyrmion-skyrmion interaction was calculated by fixing one skyrmion and changing the position of the other skyrmion, using periodic boundary conditions along both directions in the plane. From the obtained total energies we subtracted a reference spectrum for a homogeneously magnetized periodic surrounding around the skyrmion fixed at different positions in the mesh. Small deviations arising due to the finite mesh are mostly canceled out by this procedure.
	
	{\bcorr	The results for the interaction strengths are shown in Fig.~\ref{fig:11}. A smooth energy function rapidly decaying with increasing distance between skyrmion and boundary is obtained.
	For small distances between the boundary and a skyrmion or between two skyrmions, a deformation of the quasiparticles is visible in the micromagnetic simulations, also leading to a bending of the potential in Fig.~\ref{fig:11}. 
	{\bcor During the construction of the quasiparticle model it is assumed that the thermal energy $k_{\textrm{B}}T$ is lower than the energy value where the potential function starts to bend, which is around $5\,$meV ($60\,$K) for the skyrmion-skyrmion interaction potential in Fig.~\ref{fig:11}. In this regime it can be assumed that for sufficiently short simulation times the strong deformation of the skyrmions due to the interactions becomes exceedingly rare, and the basic spin structure is maintained. This temperature regime is in reasonable qualitative agreement with the range where the escape, collapse or creation of skyrmions may also be excluded, discussed in detail in Fig.~\ref{fig:3}.}
	It was shown in Ref.~\cite{LiRe2013} that the interaction potentials in this regime may be approximated by the exponential model function $E(r)=a\exp(-r/b)$. This is confirmed by the simulation results in Fig.~\ref{fig:11}, with the fitted model parameters $a = 0.211\,$meV, $b = 1.343\,$nm for skyrmion-boundary (Sk-Bnd) repulsion and $a= 1.246\,$meV, $b=1.176\,$nm for skyrmion-skyrmion (Sk-Sk) interaction. The similar characteristic length scales $b$ between the two cases support that the same physical mechanism is responsible for the interactions, namely the formation of chiral noncollinear spin structures due to the DMI as discussed {\bcor in the main text}. The different values of the parameter $a$ can be reformulated as a shift of the exponential function along the $r$ axis. While this distance is measured between two magnetization cells pointing oppositely to the homogeneous background in the case of two skyrmions, in the case of the skyrmion-boundary repulsion it is measured between the center of the skyrmion and the magnetization at the edge which is only slightly tilted from the homogeneous background, meaning that the same interaction strength is reached at a smaller distance in the latter scenario.
	\begin{figure}
		\includegraphics[width=0.9\linewidth]{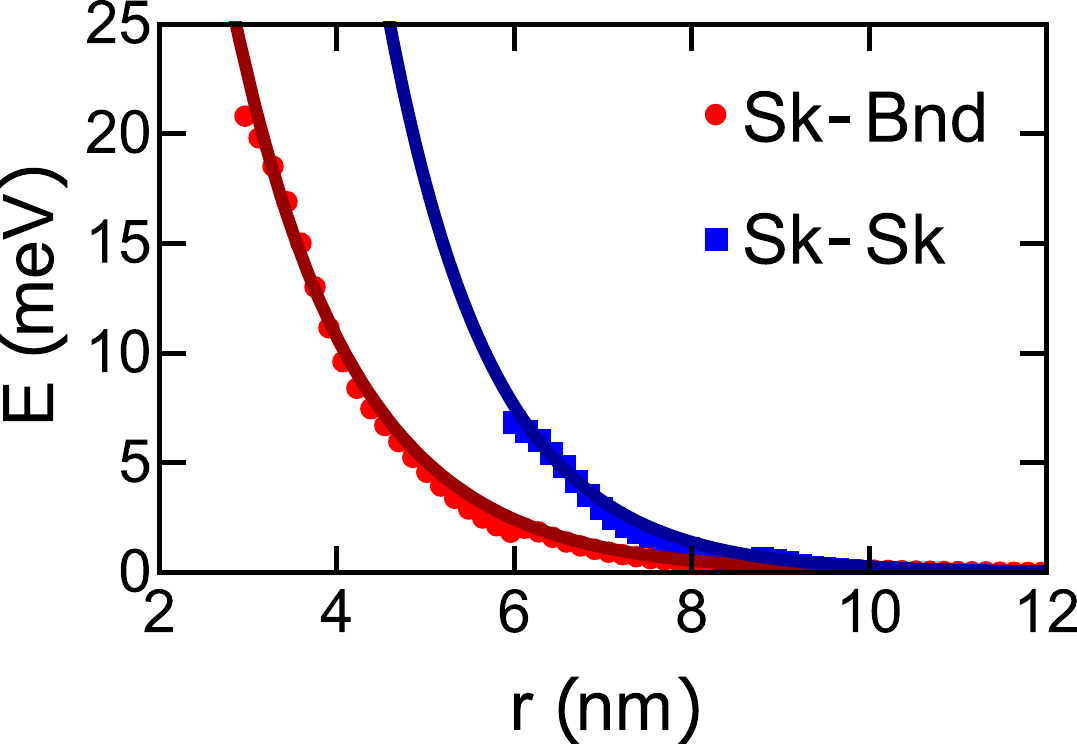}
		\caption{
			\textbf{Potential energy depending on the skyrmion-skyrmion and skyrmion-boundary distance.}
			Micromagnetic interaction energies between skyrmion and skyrmion (blue squares), and between skyrmion and boundary (red circles) on a finite nanoisland. Solid lines display exponential fits based on the model function $E(r)=a e^{-r/b}$; the fitting parameters are skyrmion-boundary (Sk-Bnd): $a = 0.211\,$meV, $b = 1.343\,$nm; skyrmion-skyrmion (Sk-Sk): $a= 1.246\,$meV, $b=1.176\,$nm.}\label{fig:11}
	\end{figure}
}

\subparagraph{Quasiparticle simulations.}
	For the calculation of the stochastic motion of skyrmions following the quasiparticle approach, the Metropolis algorithm \cite{metropolis1953equation} was utilized. With this method, the probability density function converges to the Boltzmann distribution determined by the energy functional of the system.
	For the triangular geometry, the potential surface was computed by taking the superposition of the potentials shown in Fig.~\ref{fig:11} from the three boundaries for every grid point of the chosen finite, rectangular mesh. The cell size was $\Delta V=0.5\times 0.5\times 0.4\,$nm.
	The initial positions of the skyrmions were randomly chosen inside the confined structure, excluding the case in which both quasiparticles start from the same simulation cell. 
	At each time step, a possible adjacent position is selected  for each skyrmion simultaneously via pseudo-random numbers. If the energy of the attempted new state, including interaction between the skyrmions, is lower than the energy of the initial one, the skyrmion will move there. If not, the transition into this state happens with a probability of $p(\Delta E)=\exp(-\Delta E/ k_B T)$ following the Boltzmann distribution. Here $\Delta T= E_2 - E_1$ is the energy difference between the final and the initial states, $k_B$ is Boltzmann's constant and $T$ is the temperature. 
	Subsequently, new attempted positions are generated and accepted or refused as before until a fixed number of simulation steps is reached.

\bibliographystyle{naturemag}

\section*{Acknowledgements}
Financial support was provided by the Deutsche Forschungsgemeinschaft (DFG) via CRC/TRR 227 and SFB 762, by the European Union via the Horizon 2020 research and innovation program under Grant Agreement No. 665095 (MAGicSky), by the Alexander von Humboldt Foundation, and by the National Research, Development and Innovation Office of Hungary under Project No. K115575.

\section*{Author Contributions}
A.F.S. conducted and analyzed the micromagnetic and quasiparticle Monte Carlo simulations. E.Y.V. proposed and developed a general concept of this investigation. All authors discussed the results and contributed to the manuscript.
\\

\section*{Competing Interests statement}
The authors declare no competing financial interests.

\section*{Data availability}
The code and the data that support this work's findings are available from the corresponding author on request. 

\end{document}